\documentclass[10pt,a4paper]{article}

\usepackage[latin1]{inputenc}
\usepackage{fourier}
\usepackage{graphicx}
\usepackage{amsmath}
\usepackage{amsfonts}
\usepackage[latin1]{inputenc}
\usepackage[T1]{fontenc}
\usepackage{verbatim}
\usepackage{latexsym}

\usepackage{hyperref}
\hypersetup{backref,colorlinks=true,citecolor=blue}

\title{Extracting individual contributions from their mixture: a blind 
source separation approach, with examples from space and laboratory plasmas}

\author{Thierry Dudok de Wit \\ \normalsize LPC2E, Observatoire des Sciences de l'Univers en R\'egion Centre, \\ \normalsize 3A avenue de la Recherche Scientifique, 45071 Orl\'eans cedex 2, France}

\date{\normalsize \textit{Expanded version of an article to appear in Contributions to Plasma Physics (2010), Special issue on the International Workshop on Plasma Diagnostics}}

\begin{document}
\maketitle

\begin{abstract}
Multipoint or multichannel observations in plasmas can frequently be modelled as an instantaneous mixture of contributions (waves, emissions, ...) of different origins. Recovering the individual sources from their mixture then becomes one of the key objectives. However, unless the underlying mixing processes are well known, these situations lead to heavily underdetermined problems. Blind source separation aims at disentangling such mixtures with the least possible prior information on the sources and their mixing processes. Several powerful approaches have recently been developed, which can often provide new or deeper insight into the underlying physics. 

This tutorial paper briefly discusses some possible applications of blind source separation to the field of plasma physics, in which this concept is still barely known. Two examples are given. The first shows how concurrent processes in the dynamical response of the electron temperature in a tokamak can be separated. The second example deals with solar spectral imaging in the Extreme UV and shows how empirical temperature maps can be built.

\end{abstract}

\section{Introduction}

Many physical phenomena can be described as a linear superposition of some elementary processes. In acoustics, for example, recordings made by a array of microphones can be adequately described as a linear superposition of sounds that originate from different sources. The problem of recovering these individual sources from their mixtures is a longstanding problem. A hallmark for this is the \textit{cocktail party problem} \cite{haykin05}, in which a person tries to isolate a voice that is hidden in the ambient noise. Blind Source Separation (BSS) is a recent and powerful framework for recovering the original sources from their mixtures using the least prior knowledge on the sources and on their mixing process \cite{hyvarinen01, comon10}. 

Why blind separation ? If some properties of the sources are known beforehand, then this prior information should be incorporated in the extraction and a Bayesian approach is appropriate \cite{kuruoglu10}. Likewise, if the sources are known to interact non-linearly, then nonlinear techniques should be considered, e.g. Ref. \cite{shi09}. Quite frequently, however, the sources are not known beforehand, and so an unsupervised analysis is required before we may consider a model that incorporates more information. In the last decade, BSS has become a very active field of research with applications in disciplines such as acoustics \cite{pedersen08}, computer vision \cite{hyvarinen09}, astrophysics \cite{leach08}, multispectral imaging \cite{collet10}, and many more. In plasma physics, many observations based on multichannel diagnostics are good candidates for doing BSS and yet, very few applications have bee reported so far.

This article aims at giving a short introduction to BSS in the context of multichannel or multipoint plasma diagnostics. Two examples are given: the first one is about coherent electron temperature fluctuations in a tokamak, and the second one on solar spectral imaging in the Extreme UltraViolet (EUV). In both examples, we consider multichannel data and show how these can be adequately modelled as a linear superposition of individual sources. Some comments and a conclusion follow in the last Section.


\section{How does BSS work ?}

There exist several approaches for doing BSS, depending on the assumptions one is making on the sources and on their mixing coefficients. Let us assume that we have a bivariate data set $I[x,t]$ that consists of measurements made simultaneously at discrete times $t$ and at specific locations $x$ (or any other combination of two variables, such as spectra $I[\lambda,t]$). These data are most conveniently stored in a matrix. BSS consists in decomposing the data into a separable set of modes of time and space
\begin{equation}
I[x,t] = \sum_{k=1}^N \phi_k[t] \; \psi_k[x] + \epsilon[x,t] \ ,
\label{eq:bss1}
\end{equation}
where $\epsilon[x,t]$ is an error term to be minimised. Depending on the context, the spatial modes $\psi_k[x]$ are called \textit{sources} and the temporal ones are called the \textit{mixing coefficients}, or conversely. 

\subsection{Singular Value Decomposition (SVD)}

Equation \ref{eq:bss1} is heavily underdetermined and so the temporal and spatial modes need to be constrained. The choice of the constraint leads to different variants of the BSS. A convenient constraint is orthonormality, i.e.
\begin{equation}
\sum_t \phi_k[t] \phi_l[t]  =  \sum_x \psi_k[x] \psi_l[x] = \left\{
\begin{array}{lll}
0 & \textrm{if} & k \neq l \\
1 & \textrm{if} & k = l
\end{array}
\right. \ .
\end{equation}
This requires the definition of weights $A_k$, finally to obtain
\begin{equation}
I[x,t] = \sum_{k=1}^N A_k \; \phi_k[t] \; \psi_k[x] + \epsilon[x,t] \ .
\label{eq:svd}
\end{equation}
This decomposition is unique\footnote{except when there are identical weights.} and has numerous interesting properties. In plasma physics, it goes under the name of principal component analysis \cite{chatfield95}, Singular Value Decomposition (SVD) \cite{press02}, or biorthogonal decomposition \cite{aubry95b}. Many applications to the extraction of coherent fluctuations from multichannel plasma measurements have been reported, see for example Refs. \cite{ddw94, kim99, dinklage00, carey04}.

The simplicity of the SVD is also one of its major shortcomings. Indeed, the modes $\phi_k[t]$ and $\psi_k[x]$ are nothing but the eigenvalues of the sample covariance matrix and so they rely on second order statistical properties only. For that reason, the SVD cannot properly distinguish random variables with a Gaussian probability density from those that are non-Gaussian. Quite often, this departure from Gaussianity is precisely one of the properties that gives deeper insight into the underlying physics \cite{frisch95} and so a different approach is required.

\subsection{Independent Component Analysis (ICA)}

The SVD was the classical technique for doing BSS until the  mid-90's, when new algorithms opened the way for source separation based on higher order statistical moments \cite{cardoso01}. An interesting generalisation of the SVD is  \textit{Independent Component Analysis} (ICA) \cite{hyvarinen01,comon94}, which consists in applying the model of Eq.~\ref{eq:bss1} while constraining the modes to be independent rather than orthonormal. Independence implies
\begin{equation}
\mathcal{P}(\phi_k, \phi_l) = \mathcal{P}(\phi_k) \cdot \mathcal{P}(\phi_l) \qquad \textrm{and} \qquad 
\mathcal{P}(\psi_k, \psi_l) = \mathcal{P}(\psi_k) \cdot \mathcal{P}(\psi_l)  \ ,  
\end{equation}
where $\mathcal{P}(\cdot)$ stands for the probability density. Independence is a more stringent and often a more meaningful constraint than orthogonality. For Gaussian data, the ICA and the SVD give similar results.  Non-Gaussianity, however, provides an additional leverage that is precisely used by ICA. Interestingly, ICA can be considered as a way to go against the central limit theorem: while a mixture of random variables tends to be more Gaussian than its individual constituents, ICA tries to recover the constituents that are maximally non-Gaussian. 

The computation of the SVD is straightforward whereas that of the ICA requires the maximisation of a measure of non-Gaussianity. Different algorithms have been developed, such as JADE \cite{cardoso98} and FastICA \cite{hyvarinen00}. Although many applications have been reported (see for example \cite{hyvarinen01} for a list of them), there have been only very few applications  to plasmas so far \cite{verdoolaege04, ddw07}.

\subsection{How many sources ?}

One of the crucial issues with BSS is the determination of the number $N$ of sources in Eq. \ref{eq:bss1}. Until recently, it was widely believed that this number could not exceed the number of observables, otherwise the problem would be underdetermined. Multiresolution (i.e. wavelet) techniques have opened  the way for new concepts that allow to overcome this limitation. The key concept here is sparsity, which we shall come back to in the last Section.

With the SVD, the number $N$ of sources equals the smallest dimension of the data matrix. For multichannel data, this number usually corresponds to the number of channels. The salient features are captured the modes that have large weights $A_k$ so all one has to do is identify the number of outstanding weights. In ICA, on the contrary, the number of sources must be specified beforehand. For that reason, careful validation is crucial.


\subsection{Positive source separation}

The independence constraint in the ICA is not always an appropriate condition for separating sources. There are instances where the positivity of the sources and their mixing coefficients also needs to be imposed. Consider for example soft X-ray measurements in a plasma column made by a pinhole camera. Each line-integrated measurement is a linear combination of positive contributions (there are no negative sources) that originate from different regions, with positive mixing coefficients (contributions can only be additive).  

Positive Blind Source Separation (positive-BSS) has become a very active field of research since the advent of efficient numerical schemes. A powerful method for doing positive-BSS is Bayesian Positive Source Separation (BPSS) \cite{moussaoui06}, which is the method we shall use below. There exists, however, a variety of other approaches that differ in the way the positivity constraint is applied \cite{lee99,lee01,cichocki02,nascimento05,comon10}. Although the convergence properties of positive-BSS algorithms have received a lot of attention, careful validation is  important to avoid ending up with local solutions.

By using the BPSS, we assume that the data can be decomposed as in Eq.~\ref{eq:bss1} with in addition $\phi_k[t] \ge 0$ and 
$\psi_k[x]\ge 0 \ \forall k$. The modes are stored in random matrices whose a posteriori distribution is given by Bayes' theorem
\begin{equation}
\mathcal{P} \left( \phi, \psi | I \right) = \frac{\mathcal{P} \left(I | \phi, \psi\right) \; \mathcal{P} \left( \phi, \psi \right)}{\mathcal{P} \left(I \right)} \ .
\end{equation}
We assume that the distribution functions of $\phi$ and $\psi$ are zero for negative values, which can be enforced by taking for example a Gamma distribution. For more details, see Ref. \cite{moussaoui06}.


\section{First example : temperature fluctuations in a tokamak}

Let us first consider temporal mixtures first with electron temperature fluctuations in the Tore Supra tokamak, as measured by electron cyclotron emission \cite{segui05}. We focus on the stationary regime of an ohmic discharge with the injection of a small hydrogen pellet. The temperature is recorded at 12 positions, radially distributed from $r/a = 0.1$ to $0.81$, where $a=0.70$ m is the minor radius. The dynamics of the plasma reveals two types of phenomena: the ubiquitous sawtooth instability with its periodic oscillation, and the sharp temperature drop caused by the pellet, followed by a slow relaxation. The recovery of the radial profile of the diffusive and convective transport processes from the dynamical response of the plasma is an important issue here  \cite{lopescardozo95} but it is hampered by the simultaneity of the two perturbations. Therefore, it is important to find a means for disentangling the two perturbations.

The fluctuation levels are small ($< 8 \%$) so the dynamic response of the plasma may reasonably be assumed to be linear. This response can then be expressed as linear superposition of eigenfunctions of the transport operator \cite{moret92}. We know that the two types of perturbations are independent in time, so in the following we try to separate them by using BSS. No positivity of the modes is enforced in this first example because we are interested in small fluctuations around a mean value.

A proper normalisation is required because the residual error to be minimised is scaling-dependent. For fluctuations, two reasonable normalisations are
\begin{equation} 
I[t] \longrightarrow \frac{I[t] - \langle I \rangle}{\langle I \rangle} , \qquad I[t] \longrightarrow \frac{I[t] - \langle I \rangle}{\sigma_I} \ ,
\end{equation}
where $\sigma_I$ is the standard deviation of $I[t]$ at a given location, and $\langle I \rangle$ is its time-average. The second normalisation (called standardisation in statistics) gives the same weight to all channels, so it may excessively amplify those channels that have small fluctuation levels. For that reason, we prefer the first normalisation whose result is illustrated in Fig. 1. There are 12 channels with 586 samples each, so the maximal number of sources is 12.

\begin{figure}[!htb]
\includegraphics[width=0.98\linewidth]{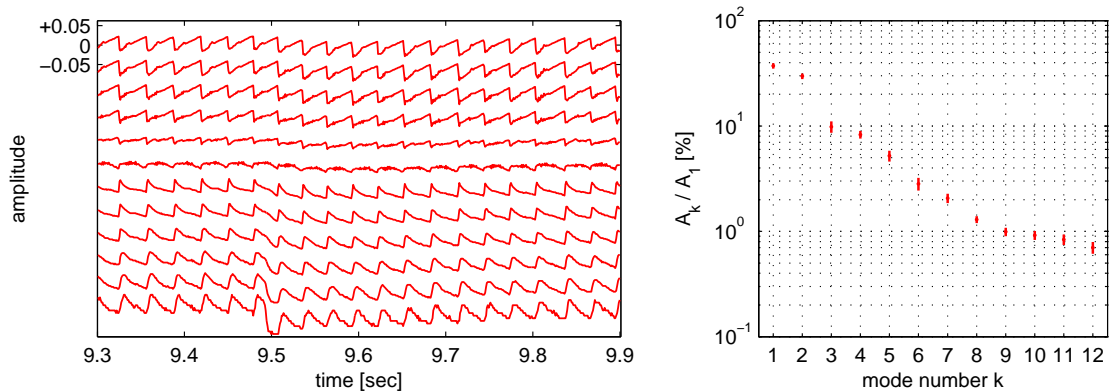}
\caption{Left plot: Relative variation of the electron temperature from the centre (top) to the edge of the plasma (bottom). The scale is the same for all time series, which have been shifted vertically for easier visualisation. Right plot: the distribution of the weights $A_k$ obtained by SVD. The 2$\sigma$ confidence intervals are obtained by bootstrapping.}
\label{fig1}
\end{figure}

The distribution of the weights $A_k$ obtained by SVD suggests that 2 to about 6 modes are required to reproduce the salient features of the dynamics, see the right plot in Fig. 1. Their inspection indeed confirms that the 5 first ones capture coherent fluctuations, whereas the remaining ones mostly describe random fluctuations. To apply the ICA, we first prewhiten the data by computing the $N$ largest SVD modes, and then apply ICA to these modes rather than to the data. The results are given in Fig.~2.

\begin{figure}[!htb]
\includegraphics[width=0.98\linewidth]{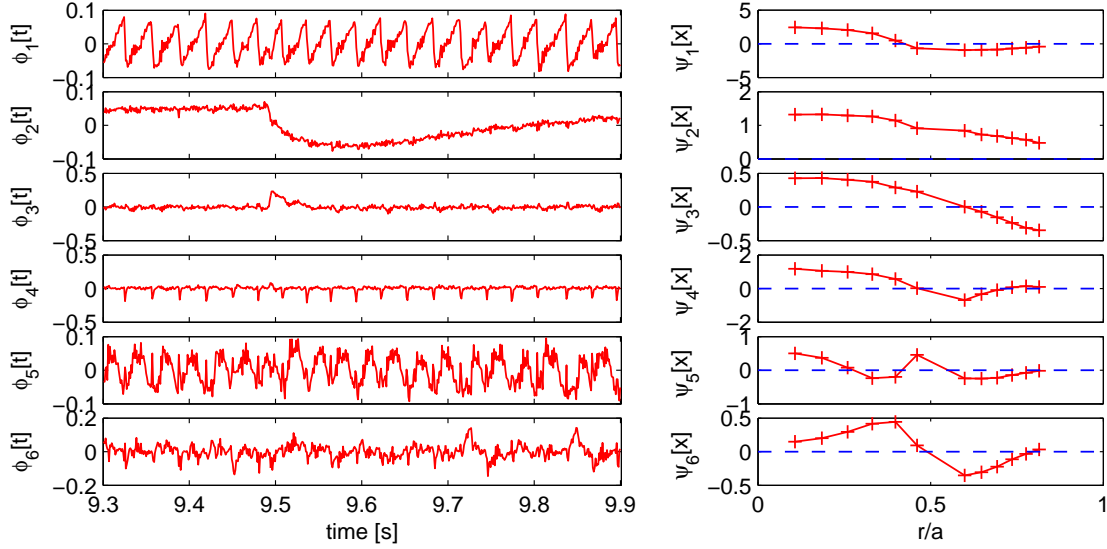}
\caption{The six temporal modes $\phi_k[t]$ obtained by ICA (left column) and their corresponding spatial modes $\psi_k[r/a]$ (right column). All the temporal modes have unit norm.}
\label{fig1}
\end{figure}

The key result is that modes 2 and 3 capture almost exclusively the signature of the pellet whereas the other ones exhibit sawtooth activity only; it is remarkable that a statistical method like ICA succeeds so well in separating so well two different physical processes. Such a separation cannot be achieved by SVD, which mixes the modes (not shown).

One can now reconstruct the dynamic response to the sawtooth instability only by summing over the corresponding modes ($I_{saw}[t,x] = \sum_{k=1,4,5,6} \phi_k[t] \psi_k[x]$) whereas the response of the plasma to the pellet can be obtained by summing over the second and third modes only. Note, however, that the separation is not total as modes $\phi_1[t]$ and $\phi_4[t]$ still carry a weak signature of the pellet injection. Nevertheless, the quality of the reconstruction is largely sufficient to carry out a perturbative transport analysis. 

Concerning the choice of the number of sources, we find that modes 1 to 5 remain unaffected when $N=5$ or more sources are selected, or when standardised data are used. Modes 6 and beyond capture mostly incoherent fluctuations that cannot be ascribed to known physical processes, see Fig. 2. These are strong indications that $N=5$ sources are required to properly describe the dynamics of the electron temperature. More sources might be needed with a lower noise level or with more ECE channels.


\section{Second example : solar spectral imaging in the EUV}

In this second example, we consider spatial mixtures with spectroscopic data from the Sun and use positive-BSS. The solar spectral irradiance in the EUV results from the superposition of emissions that are caused by different physical processes in the highly complex and dynamic solar atmosphere \cite{phillips08}. There is strong observational evidence for the spectral variability to arise from a linear combination of a few elementary spectra with different time evolutions \cite{lean82}.  This remarkable spatial coherency of the spectral variability is rooted in the strong structuring of the solar atmosphere by the intense magnetic field. In \cite{amblard08} we studied the spectral unmixing of solar EUV spectra. Here we concentrate on the spatial unmixing of solar images taken in the EUV.


\begin{figure}[!htb]
\includegraphics[width=0.9\linewidth]{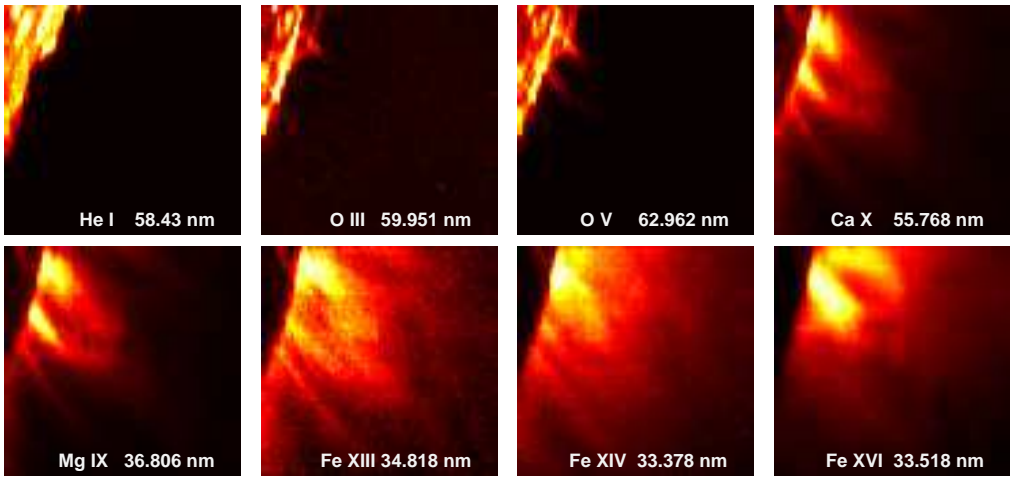}
\caption{Raster of an active region at the west limb taken on September 6, 1996, at 6:24:33 UT by the CDS instrument onboard the SoHO spacecraft; 8 emission lines out of 14 are shown. The characteristic temperature of the lines increases logarithmically from top left (20,000 K) to bottom right (2.5 MK). The spectral lines are indicated on each image. A linear vertical scale is used for all images.}
\label{fig3}
\end{figure}

Figure~\ref{fig3} shows a series of 2D images of the solar limb, taken by the Coronal Diagnostic Spectrometer (CDS) onboard the SoHO satellite \cite{brekke00}. In this example, CDS was used to measure the intensity of 14 spectral lines, with a spectral resolution better than 0.06 nm. The intensity of each spectral line depends on various plasma parameters, and in particular on the temperature. A quantitative picture of the temperature distribution, however, can only be obtained by time-consuming comparisons with simulations from radiation transfer models, and at the price of several assumptions. There is a clear need here for finding new and more empirical means to rapidly infer pertinent physical properties from such data cubes.

A conspicuous feature in Fig.~\ref{fig3} is the similarity between the different images. There are two reasons for this. First, the plasma can be multithermal so that the same line of sight (i.e. pixel) has contributions in different spectral lines. Second, the temperature response associated with each spectral line is generally wide, and sometimes even multimodal.
One can therefore assume that the intensity of each pixel to be a linear mixture of a few elementary sources, whose structure we'll now extract. 

Each CDS image is $85 \times 87$ pixels in size. We unfold the $85 \times 87 \times 14$ data cube into a $7395 \times 14$ matrix by lexicographically ordering each image. In doing so, we implicitly assume that the pixel intensities $I[x, \lambda]$ can be expressed as a separable set of spatial and wavelength components, as in Eq.~\ref{eq:bss1}.

The SVD and ICA are not applicable here as both the sources and their mixing coefficients need be positive. We consider BPSS instead, and assume that the sources$\{ \psi_k[x] \}$ and the mixing coefficients $\{ \phi_k[\lambda] \}$ are random matrices whose elements are independent and distributed according to Gamma probability density functions. In what follows, each image is normalised to its mean intensity and the sources are normalised to have unit norm.

An inspection by SVD suggests that there are 2 to 6 significant sources out of 14. From the analysis of the root mean squared error of the difference between the original data and the intensities reconstructed from the positive sources, this number should be between 3 and 5. An inspection of the sources for different solar events, however, shows that when more than 3 sources are selected, the 3 first ones remain almost unchanged, whereas the subsequent ones are smaller in  intensity (when the mixing coefficients are normalised to unit norm) and are event-dependent. These results suggest that $N=3$ sources should be considered, and that additional terms should rather be considered as minor corrections.

\begin{figure}[!htb]
\includegraphics[width=0.75\linewidth]{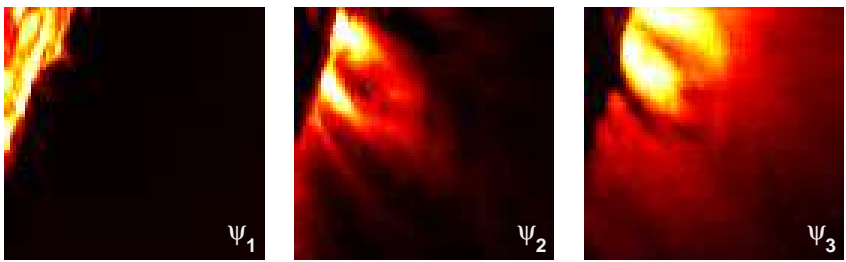}
\caption{The three spatial sources obtained by BPSS.}
\label{fig4}
\end{figure}

\begin{figure}[!htb]
\includegraphics[width=0.50\linewidth]{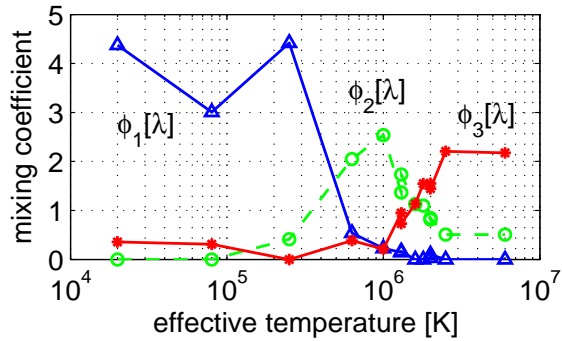}
\caption{Mixing coefficients associated with the sources shown in Fig.~\ref{fig4}, plotted versus the characteristic emission temperature of each spectral line.}
\label{fig4b}
\end{figure}

Figure~\ref{fig4} shows the 3 sources obtained by BPSS and their mixing coefficients. In contrast to the SVD and the ICA, the BPSS reveals structures that can be readily linked to known processes in the solar atmosphere. Interestingly, the mixing coefficient reveal a clear temperature ordering as the 3 sources capture emissions that respectively originate from cool (1), medium hot (2) and hot (3) plasmas. To support this, we have plotted in Fig.~\ref{fig4b} the amplitude of the mixing coefficients versus the characteristic temperature at which the emission of each spectral line peaks. The correspondence between a spectral line and a given temperature is of course very approximate but this plot nevertheless shows a clear temperature structuring in the mixing coefficients. A quantitative comparison with the more meaningful differential emission measure is under way.

In Fig.~\ref{fig4}, the coolest source is associated with the chromosphere, in which the quiet Sun stands out as a bright disk. Two groups of magnetic loops appear at the limb; note in particular how they are entangled. The key result is that the emission in all 14 spectral lines, in spite of the complexity of the underlying atomic physics, can be expressed in terms of 3 temperature bands only, whose properties can be inferred from the data without imposing a physical model. These results are robust, in the sense that the same temperature ordering is obtained for other solar regions or for different events, provided that all three temperature bands are properly represented in the observations.

\begin{figure}[!htb]
\includegraphics[width=0.5\linewidth]{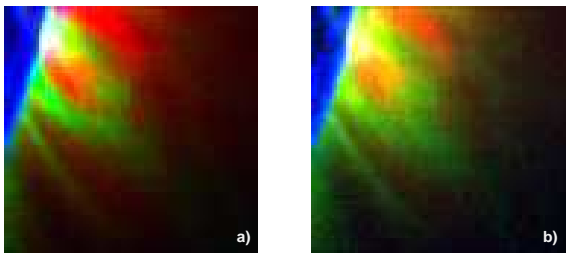}
\caption{Multispectral representation of the solar limb in which source 3 is assigned to the red channel, source 2 to the green channel and source 1 to the blue channel (a). Figure (b) has been obtained in a similar way by assigning the intensity of the Fe XIII, Si XII and He I lines respectively to the red, green and blue channels.}
\label{fig5}
\end{figure}

We can now make a compact representation of the solar corona by compressing all 14 images into one single colour plot. We do so by assigning the hot, medium hot and cool sources respectively to the red, green and blue channels. The result is displayed in Fig.~\ref{fig5}. Also shown is the same picture obtained by assigning the coronal Fe XIII and Si X lines, and the chromospheric He I line respectively to the red, green and blue channels. Although the two pictures look similar, the former has a sharper contrast because of the better separation of the temperature. This allows the loop structure to be better seen; furthermore, the whitish pixels can now be interpreted as multithermal regions.

Such multichannel representations are useful as quick-looks but also be used as a guide for supervised classification. They are particularly useful for visualising the large stream of high resolution (4k$\times$4k) images from the recently launched SDO satellite, which continuously monitors the Sun in up to 10 spectral lines, several times per minute. Since the sources are recovered simply by linear combination of the observed images, using coefficients that are computed once for ever, the decomposition can be carried out in real-time.

\section{Outlooks and Conclusion}

The two examples we have discussed (electron temperature fluctuations in a tokamak and solar spectral imaging) both show that BSS, even though it is an empirical method, can give better insight into the physics by separating different processes. Let us stress again, however, that BSS makes sense only if we know that the dynamics can be described by a linear superposition of various contributions, so that the salient features can be captured by a few sources. 

The instantaneity of the mixture is a condition that can be partly relaxed, and some recent BSS techniques can handle convolutive mixtures \cite{comon10}. In the first example, the mixture actually was not instantaneous since the temperature perturbation diffuses through the plasma column. The propagation, however, was sufficiently fast to allow for its description in terms of a few sources.  The instantaneity constraint becomes a major obstacle when dealing with wavefields consisting of structures that move randomly or that are convected. For such wavefields, the separable model Eq.~\ref{eq:bss1} is not appropriate and indeed none of the BSS methods will properly work. If the wavefield is homogeneous, then the SVD will simply yield Fourier modes. The only exception occurs with travelling waves, which can be described 	as spatio-temporal symmetries \cite{aubry95b}.

BSS can not only be used for physical interpretation but also for prepocessing. A typical example is the tomographic reconstruction of density profiles in a plasma column. Most tomographic techniques involve a regularisation step based on concepts such as entropy maximisation, which require positive definite data. The processing of a full discharge then requires as many tomographic reconstructions as there are observations, which can become very time-consuming. The computational burden can be considerably alleviated by first estimating the positive sources, then applying the tomographic reconstruction to these few sources only, and finally combining these inverted sources to obtain the full data set.

Careful validation is a crucial issue with BSS. One should always compute the solutions for different numbers of sources and investigate the sources and their mixing coefficients before making decisions. Preprocessing also matters since these techniques are scaling dependent. In practice, one should consider a normalisation that gives equal weight to all channels or, better, that gives the same noise level to all channels. The information about the noise can actually be incorporated in the Bayesian priors, and so for better performance the techniques (in particular BPSS) should be fine-tuned to the data under consideration.

BSS is a rapidly evolving field and although recent techniques such as ICA and BPSS have already found their way into applications, much more is to come. A key concept is sparsity, which means that the data, when projected on a suitable set of basis functions (typically wavelets), require a limited amount of information to describe them \cite{mallat08}.  This implies that sources with some quantitatively measurable diversity should have differing signatures when projected on such bases, which then provide additional leverage for separating them.  Recent techniques such as morphological component analysis \cite{bobin08} take advantage of this sparsity and of the different morphological properties of the sources to do BSS. One of their assets is their ability to extract more sources than there are observables. 


\subsection*{}
I gratefully acknowledge discussions with S. Moussaoui, J. Aboudarham, P.-O. Amblard, F. Auch\`ere, G. Cessateur,  M. Kretzschmar and J. Lilensten as well as the International Space Science Institute (ISSI, Bern) for hospitality. Special thanks are due to the instrumental teams that provided the data: Tore Supra (CEA, Cadarache) and SoHO/CDS. This study received funding from the European Community's Seventh Framework Programme (FP7/2007-2013) under the grant agreement nr. 218816 (SOTERIA project, www.soteria-space.eu).


\providecommand{\WileyBibTextsc}{}
\let\textsc\WileyBibTextsc
\providecommand{\othercit}{}
\providecommand{\jr}[1]{#1}
\providecommand{\etal}{~et~al.}

\end{document}